\documentclass[sn-mathphys]{sn-jnl}

\usepackage{enumitem}

\begin{document}

\title[TAP-RAF]{Emergence of Autocatalytic Sets in a Simple Model of Technological Evolution}

\author*[1]{\fnm{Wim} \sur{Hordijk}}\email{wim@WorldWideWanderings.net}
\author[2]{\fnm{Stuart} \sur{Kauffman}}\email{stukauffman@gmail.com}

\affil[1]{\orgname{SmartAnalytiX}, \orgaddress{\city{Vienna}, \country{Austria}}}
\affil[2]{\orgname{Institute for Systems Biology}, \orgaddress{\city{Seattle}, \state{WA}, \country{USA}}}

\abstract{Two alternative views of an economy are combined and studied. The first view is that of technological evolution as a process of combinatorial innovation. Recently a simple mathematical model (TAP) was introduced to study such a combinatorial process. The second view is that of a network of production functions forming an autocatalytic set. Autocatalytic (RAF) sets have been studied extensively in the context of chemical reaction networks.

Here, we combine the two models (TAP and RAF) and show that they are compatible. In particular, it is shown that production function networks resulting from the combinatorial TAP model have a high probability of containing autocatalytic (RAF) sets. We also study the size distribution and robustness of such ``economic autocatalytic sets'', and compare our results with those from the chemical context. These initial results strongly support earlier claims that the economy can indeed be seen as an autocatalytic set.}


\maketitle

\section{Introduction}

Technological and cultural evolution are driven by a process of combinatorial innovation \cite{Arthur:09,Enquist:11}. Recently, a new mathematical model was introduced to describe and study such a process \cite{Steel:20}. This model, referred to as the {\it theory of the adjacent possible} (TAP), accurately reproduces the well-known ``hockey stick'' phenomenon observed in economic growth: a long period of slow growth followed by explosive growth in just a short time \cite{Roser:13,Karakas:19}.

The TAP model is based on the assumption that a simple cumulative combinatorial process underlies this pattern of technological and cultural evolution. At its core is the following equation:
\begin{equation}
\label{meq}
M_{t+1}  = M_t  + \sum_{i=1}^{M_t} \alpha_i \binom{M_t}{i},
\end{equation}
where $M_t$ is the number of different types of goods in an economy at time $t$, and $\alpha_i$ is a decreasing sequence of probabilities (i.e., real numbers between 0.0 and 1.0).

The main idea of this equation is that new goods (or tools, or artefacts) are created by combining any number $i$ of already existing goods. Arbitrary combinations of $i$ existing goods have a small probability $\alpha_i$ of resulting in a useful new one. The TAP model was recently studied both theoretically and with computer simulations \cite{Steel:20}. Note that this equation represents a deterministic version that does not guarantee $M_t$ to be an integer value, and it only serves to convey the general idea behind the model. A stochastic implementation that does guarantee integer values is presented below.

Alternatively, it has been suggested that an economy can be seen as an instance of an autocatalytic set \cite{Kauffman:11,Hordijk:13,Gatti:20,Koppl:22}. The notion of an {\it autocatalytic set} originally comes from chemistry, where it is informally defined as a chemical reaction network in which the molecules mutually catalyze (i.e., speed up, or facilitate) each other's formation, and which can sustain itself from a given set of basic building blocks, or ``food set'' \cite{Kauffman:71,Kauffman:86}. This notion was formalized and studied in more detail as RAF sets \cite{Hordijk:04,Hordijk:17}, and has also been implemented with real molecules in the lab \cite{Ashkenasy:04,Vaidya:12,Arsene:18,Miras:20} and shown to exist in the metabolic networks of prokaryotes \cite{Sousa:15,Xavier:20}.

In an economy, certain goods can also act as ``catalysts'' in that they facilitate the production of other goods. Examples are hammers, conveyor belts, and computers. These goods are not used up in a production process, but facilitate the production of other goods such as tables, cars, or animated movies. Yet, such catalysts are themselves products of the same economy. As such, an economy can be seen as a production function network in which the goods mutually catalyze each other's formation, sustained by a basic food set of raw materials. In other words, an economic autocatalytic set.

An interesting question is whether these two alternative views of an economy, as a process of combinatorial innovation on the one hand and as an autocatalytic set on the other, are somehow compatible. We show that the answer to this question is a resounding {\it Yes}. Using the original TAP model, we extend it with the assignment of (random) catalysis, as was hinted at recently \cite{Kauffman:21}. We then show that autocatalytic (RAF) sets have a high probability of emerging in production function networks that result from a process of combinatorial innovation. Additionally, we also study the size distribution and robustness of these economic RAF sets.

\section{Methods}

An implementation of a stochastic discrete-time version of the TAP model based on an earlier version \cite{Steel:20} is used here, with $\alpha_i = \alpha^i$ (i.e., $\alpha$ to the power $i$, for some given value of $\alpha$). The creation of a new good from a combination of already existing ``parent'' goods is then interpreted as the introduction of a new production function, where the parent goods are the inputs and the new good is the output. In addition, each newly created good is assigned as a catalyst to the already existing production functions with a given catalysis probability $p_c$. Similarly, a new production function is catalyzed by any of the already existing goods also with probability $p_c$. The resulting model is implemented as described in Algorithm \ref{algo}.

\begin{algorithm}
\caption{TAP with catalysis}\label{algo}
\begin{algorithmic}[1]
\Require $M_0, K, \alpha, {\bf M}, p_c$
\State $t \leftarrow 0$
\State Create $M_0$ initial goods labeled $1,\ldots,M_0$
\While {$M_t < {\bf M}$}
  \State $t \leftarrow t+1$; $M_t \leftarrow M_{t-1}$
  \For {$i=1,\ldots,K$}
    \State $s_i \leftarrow \alpha^i \times \binom{M_t}{i}$
    \State $r_i \leftarrow$ Poisson$(s_i)$
    \For {$j=1,\ldots,r_i$}
      \State Create a new good $x$ labeled $M_t + 1$
      \State Select $i$ random ``parents'' for $x$ from $1,\ldots,M_{t-1}$
      \State $M_t \leftarrow M_t + 1$
      \For {$y=1,\ldots,M_t$}
        \State With probability $p_c$ assign $x$ as catalyst to production of $y$
        \State With probability $p_c$ assign $y$ as catalyst to production of $x$
      \EndFor
    \EndFor  
  \EndFor
\EndWhile
\end{algorithmic}
\end{algorithm}

Note that an upper limit $K$ on the number of parents is set for numerical and computational reasons. It was already shown earlier that this does not significantly affect the overall behavior of the TAP model \cite{Steel:20}.

At each time step $t$ the $M_t$ existing goods thus form a production function network with a food set consisting of the $M_0$ initial goods, where each production function consists of a good being produced from its parents. In addition, the goods catalyze each other's production according to the catalysis probability $p_c$. Note that each production function can thus have none, one, or multiple catalysts, depending on the random catalysis assignments (with probability $p_c$). Similarly, each good may catalyze none, one, or multiple production functions.

\begin{figure}[htb]
\centering
\includegraphics[scale=0.4]{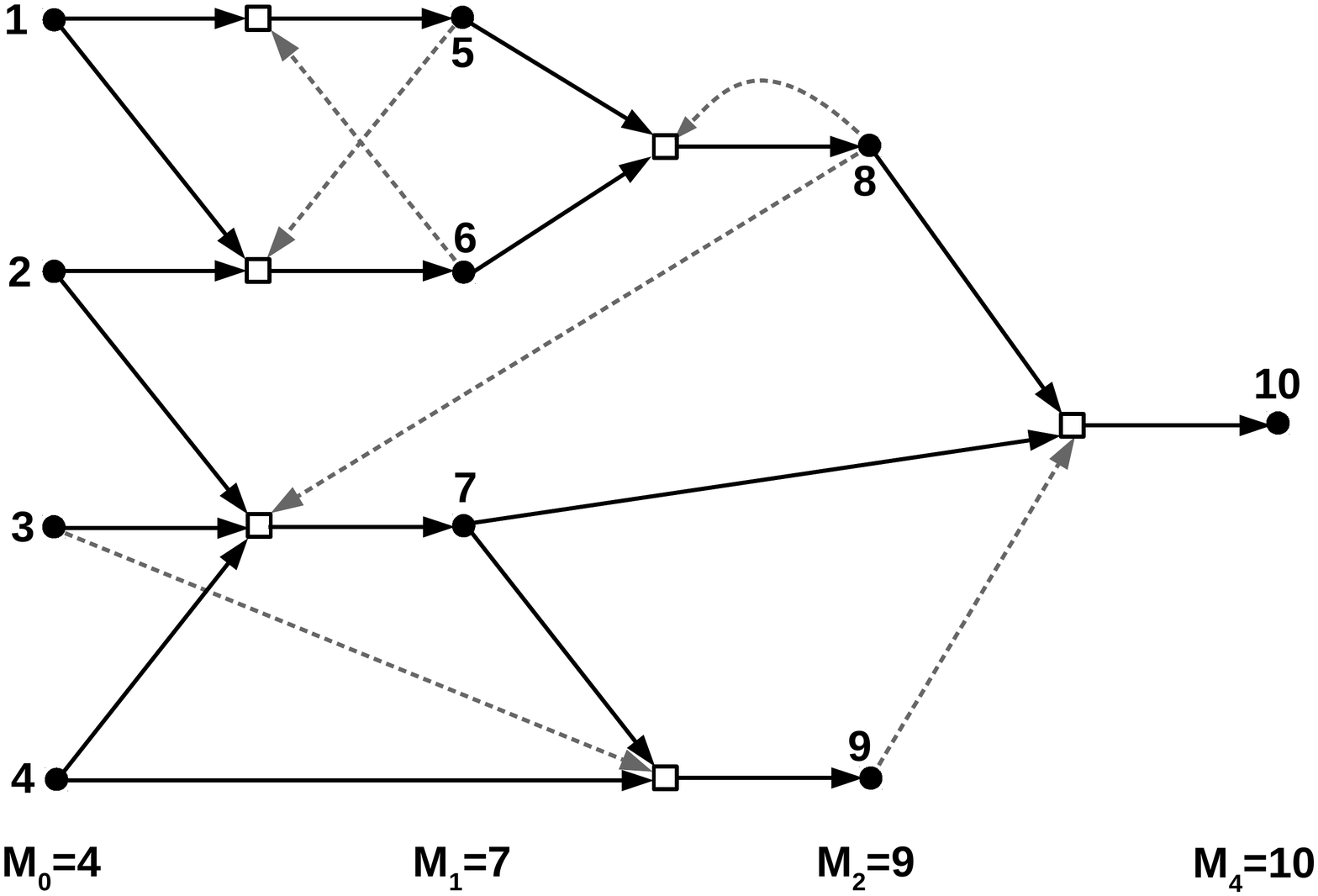}
\caption{An example production function network resulting from a TAP process, with random catalysis assignments (dashed arrows). This network also happens to form a RAF.}
\label{fig:example}
\end{figure}

Figure \ref{fig:example} shows a simple example of such a production function network, resulting from a TAP process, in a graph representation. The black dots represent the different types of goods created at each time step, and the white boxes represent their production functions. Solid arrows indicate inputs and outputs of production functions. Dashed grey arrows indicate which goods catalyze which production functions (assigned randomly).

Note that this graph representation of a production function network is similar to that of a chemical reaction network, where dots represent molecule types and boxes represent chemical reactions \cite{Temkin:96}, and with catalysis added as dashed grey arrows. Therefore, the standard RAF algorithm \cite{Hordijk:04,Hordijk:15} can be applied to such a production function network to see if an autocatalytic set (i.e., RAF set) is contained within it or not. In fact, the example network shown in Figure \ref{fig:example} constitutes a RAF: all production functions are catalyzed by some good in the set, and all goods can be produced starting from the food set, which consists of the four original goods at $t=0$.

The RAF algorithm actually finds the {\it largest} RAF (maxRAF) present in the network, which in the example above is the entire network itself. However, such a maxRAF could consist of the union of several smaller RAFs (subRAFs), including minimal, or {\it irreducible}, RAFs (irrRAFs). Indeed, there are several (smaller) subRAFs within the maxRAF of Figure \ref{fig:example}. Such subRAFs and irrRAFs can be identified by (repeatedly) applying the RAF algorithm after removing one or more (random) elements from the maxRAF.

Using the TAP simulation model including catalysis assignments, as described above, many runs are performed using different catalysis probabilities $p_c$. The RAF algorithm is then applied to the production function network resulting from these TAP processes to see how often a RAF set emerges, and how large they are. In the results presented below, most TAP model parameters are fixed as follows:
\begin{itemize}
  \item $M_0 = 10$
  \item $K = 4$
  \item $\alpha = 0.01$
  \item ${\bf M} = 1000$.
\end{itemize}
Note that rather than running a TAP simulation for a fixed number of time steps, a run will end once a threshold of ${\bf M} = 1000$ goods is crossed. So, different runs might terminate at different time steps, and result in a different number of final goods $M_t$ (but with $M_t \geq 1000$), due to the stochastic nature of the simulations. The catalysis probability $p_c$ is then allowed to vary between batches of runs, to see how it influences the possible emergence and sizes of RAF sets.

\section{Results \& Discussion}

\subsection{Probability of RAFs}

\begin{figure}[htb]
\centering
\includegraphics[scale=0.4]{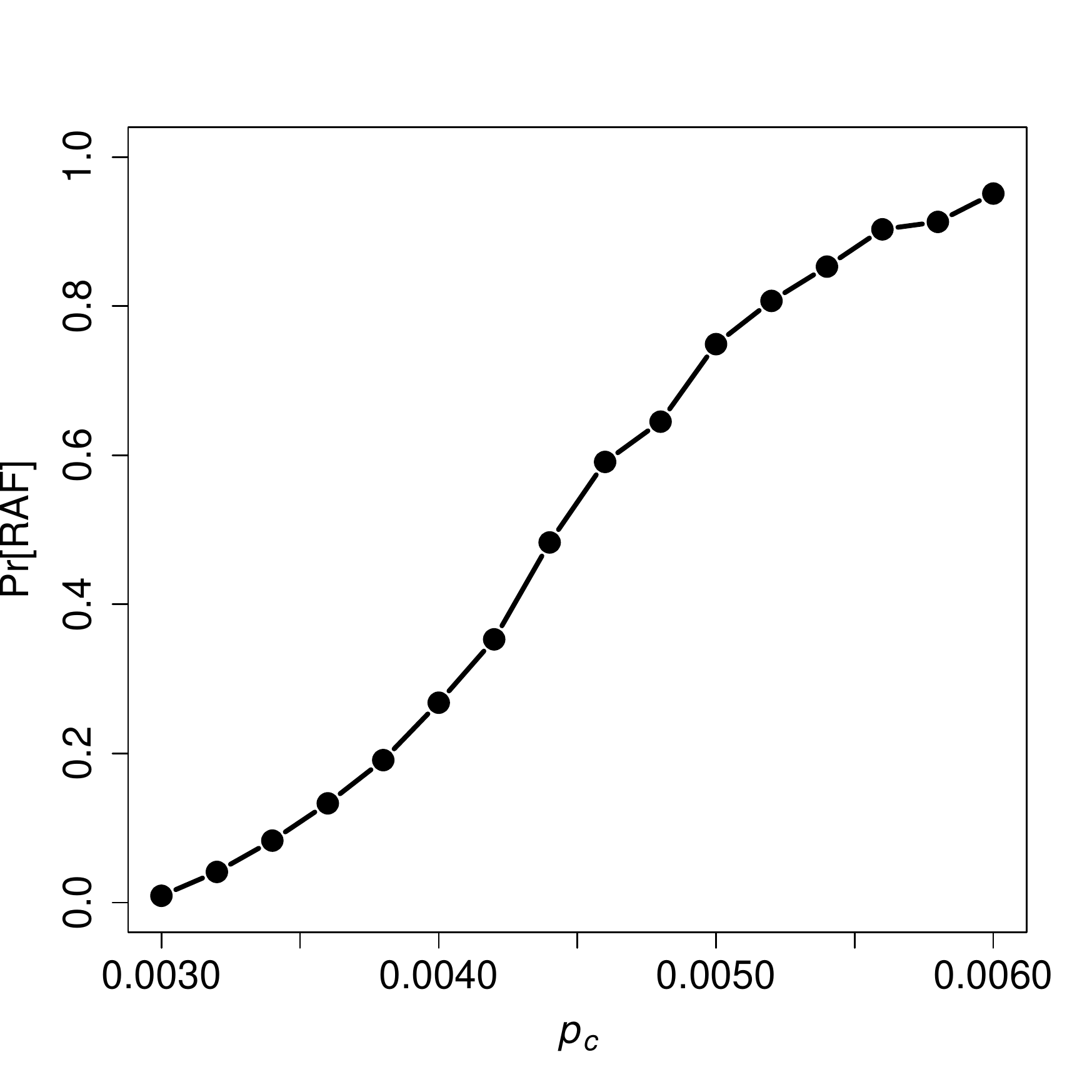}
\caption{The probability of RAFs (Pr[RAF]) for different values of the catalysis probability $p_c$.}
\label{fig:pr_RAF}
\end{figure}

First, the probability of finding RAFs at the end of a TAP run (i.e., when at least 1000 goods have been produced) is investigated. The catalysis probability $p_c$ was increased from $p_c=0.0030$ to $p_c=0.0060$ with increments of $0.0002$ (16 different values). For each value of $p_c$, 1000 runs of the TAP simulation were performed. Figure \ref{fig:pr_RAF} shows the probability of finding a (max)RAF in the final step of a TAP run (i.e., the fraction of the 1000 runs that contained a RAF) for each value of $p_c$.

The figure shows a typical S-shaped curve, going from never finding a RAF for $p_c < 0.0030$ to almost always finding one for $p_c > 0.0060$. For a value of $p_c=0.0044$ the probability of finding a RAF is close to 0.5 (i.e., about half of the runs result in a RAF). This S-shaped curve is similar to what is observed in the standard binary polymer model used in previous studies on RAFs in a chemical context \cite{Hordijk:04}. However, a higher level of catalysis is required in the TAP model to find RAFs. For example, in the binary polymer model it suffices to have each polymer catalyze (on average) between one and two chemical reactions to get a probability of RAFs around 0.5 (at least for model instances with a maximum polymer length of up to 50) \cite{Hordijk:04}. But in the TAP model it requires (on average) between five and six production functions catalyzed per good. The average size of the production function networks at the end of the runs is about $\overline{M_t}=1250$, so the average number of production functions catalyzed per good is $p_c \cdot \overline{M_t} = 0.0044 \cdot 1250 = 5.5$ to get Pr[RAF]=0.5.

An important difference between these two models is that in the binary polymer model the ratio between the number of reactions and the number of molecules grows linearly with $n$ (with $n$ being the length of the largest polymers), whereas in the TAP model the ratio between the number of production functions and the number of goods is constant (in particular, the number of production functions is exactly the same as the number of goods). The reason that there are roughly $n$ time more reactions than molecules in the binary polymer model is that each polymer of a given length $n$ can be produced from $n-1$ ligation reactions between smaller polymers \cite{Kauffman:86,Hordijk:04}.

\subsection{RAF sizes}

\begin{figure}[htb]
\centering
\includegraphics[scale=0.32]{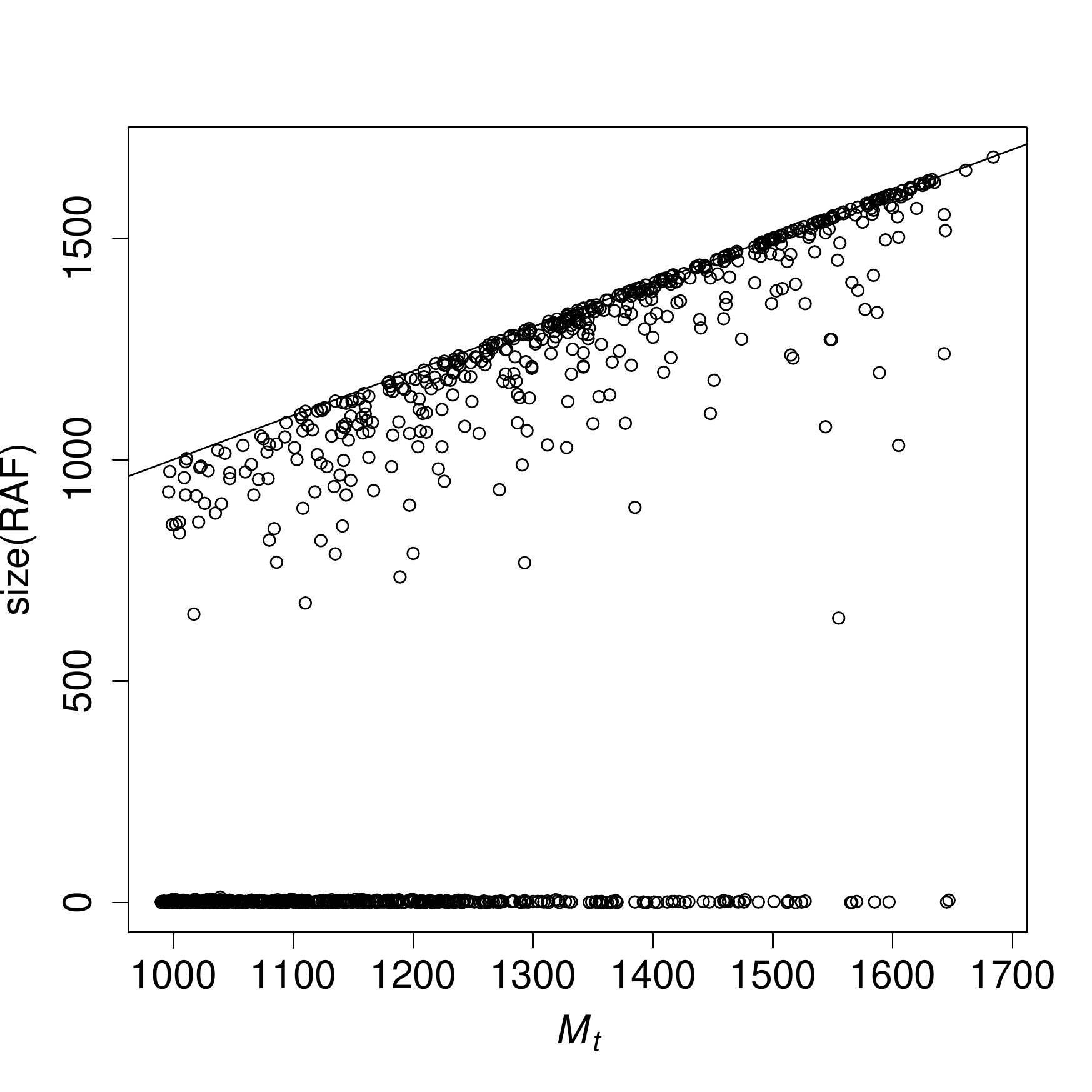}
\includegraphics[scale=0.32]{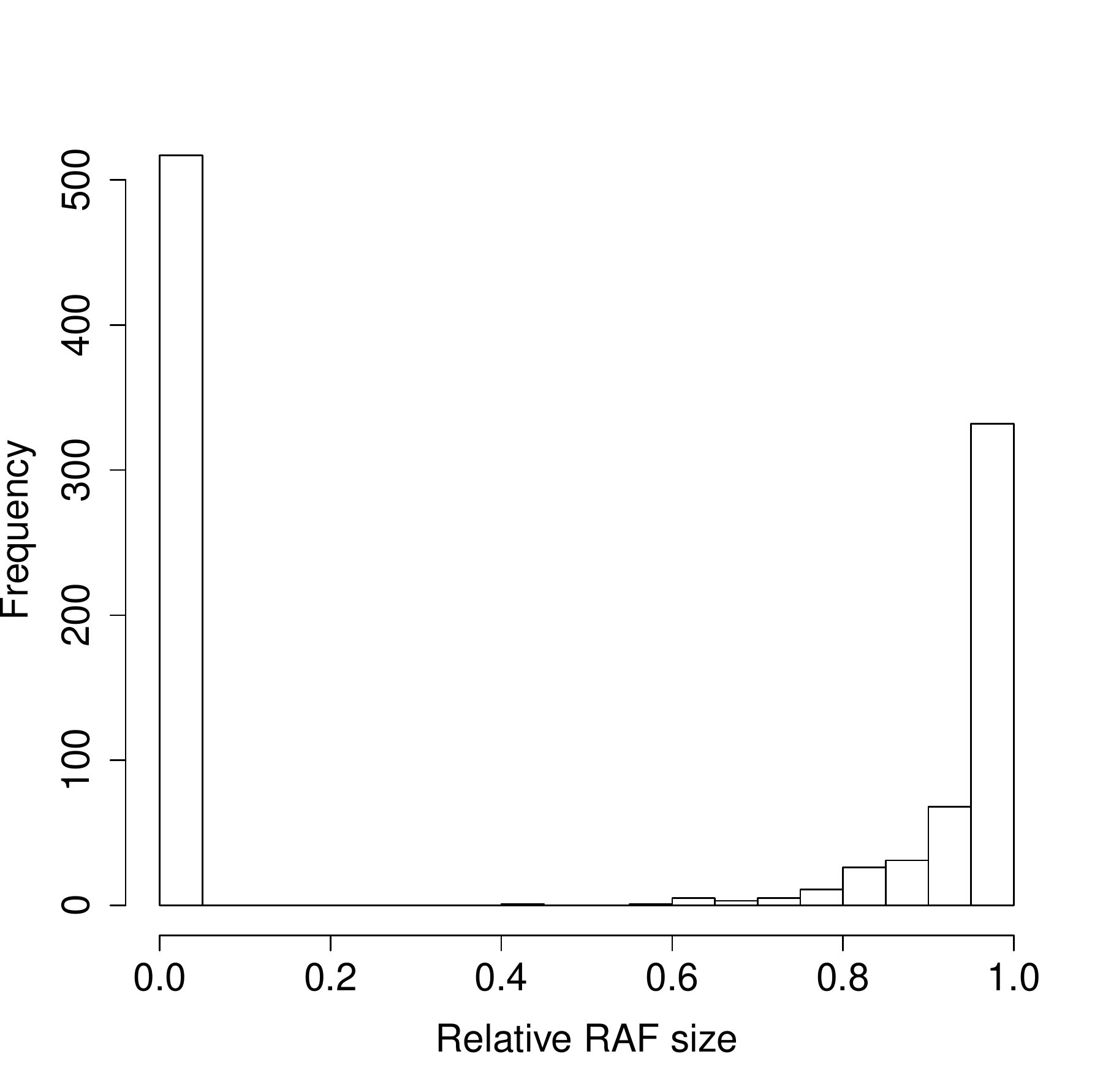}
\caption{Left: The size of a (max)RAF against the size of the full network it is part of. Right: A histogram of the relative RAF sizes (as a fraction of the full network).}
\label{fig:RAF_size}
\end{figure}

Next, the sizes of the RAFs are investigated. Figure \ref{fig:RAF_size} (left) shows a scatter plot of the maxRAF sizes (vertical axis) against the size of the full production function network (i.e., the number of goods $M_t$ at the end of the run) for the 1000 runs with $p_c=0.0044$ (which resulted in Pr[RAF]$\approx 0.5$). The line of dots at the bottom of the graph represents the close to 500 runs that did not result in a RAF (i.e., the RAF size is zero), or where there may have been a small number of production functions where one of the $M_0$ initial goods is transformed into a new good, also catalyzed by one of the $M_0$ initial goods. In the current implementation the initial items (i.e., those in the food set) were allowed to be catalysts as well (see below for more on this).

In those roughly 500 cases where there was a ``real'' RAF, the maxRAF size is anywhere from about half the total number of items $M_t$ up to the size of the full reaction network (i.e., size(RAF) = $M_t$). The far majority of RAFs are actually close to the full network in size. Figure \ref{fig:RAF_size} (right) shows a histogram of the relative maxRAF size, i.e., the RAF size divided by $M_t$. This histogram again shows the close to 500 instances with no RAF (or just a few production functions with all inputs and at least one catalyst in the food set), and around 330 instances where the RAF size is close to or equal to the full network. In other words, if a RAF exists, it immediately tends to be quite large, often consisting of the entire network. This is unlike the binary polymer model, where RAFs (when they occur with a probability of about 0.5) are usually only around 10\% of the full network (in terms of number of reactions). But note again that in the binary polymer model there is redundancy in the reactions in the sense that multiple reactions will have the same product, which is not the case in production function networks resulting from the TAP model.

\begin{figure}[htb]
\centering
\includegraphics[scale=0.32]{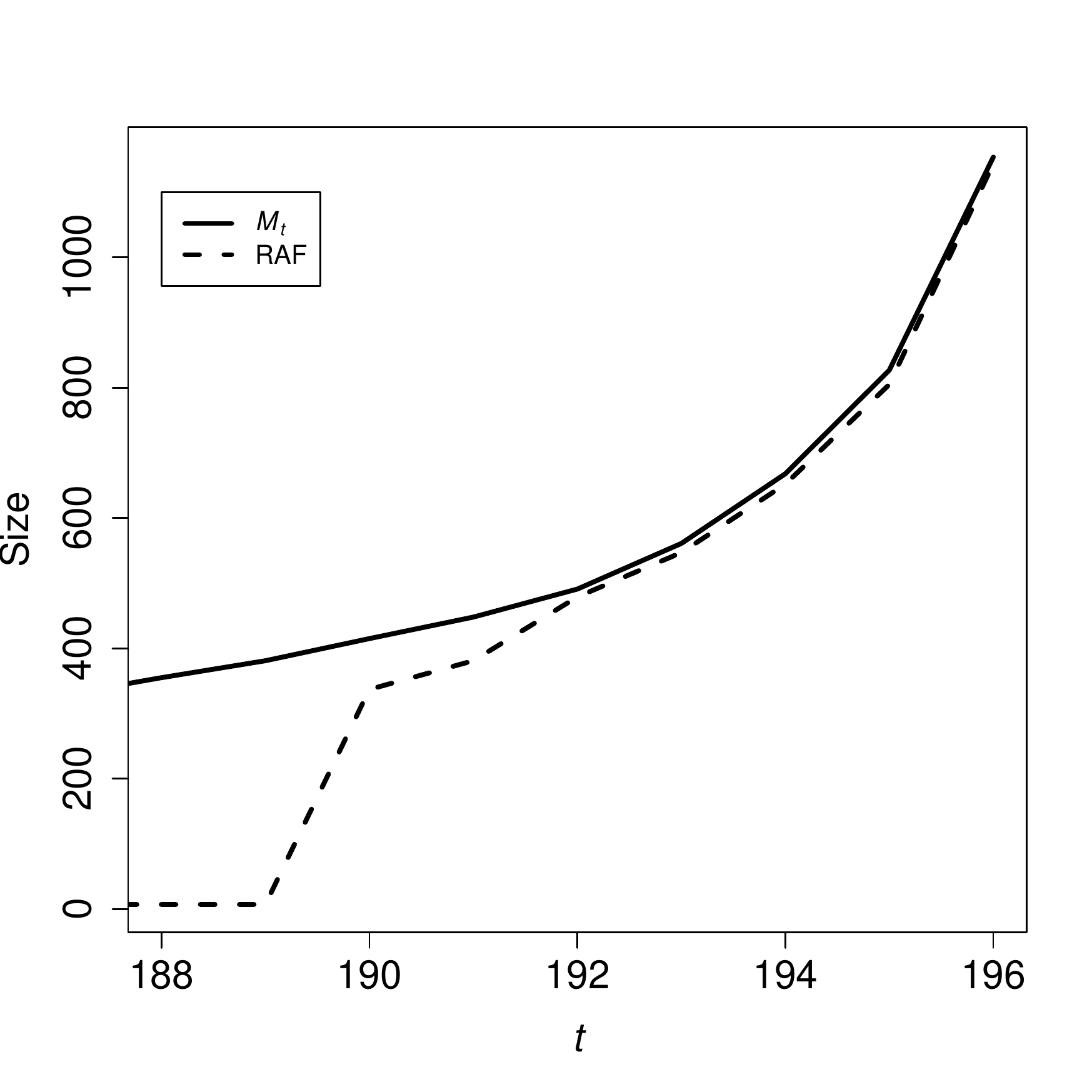}
\includegraphics[scale=0.32]{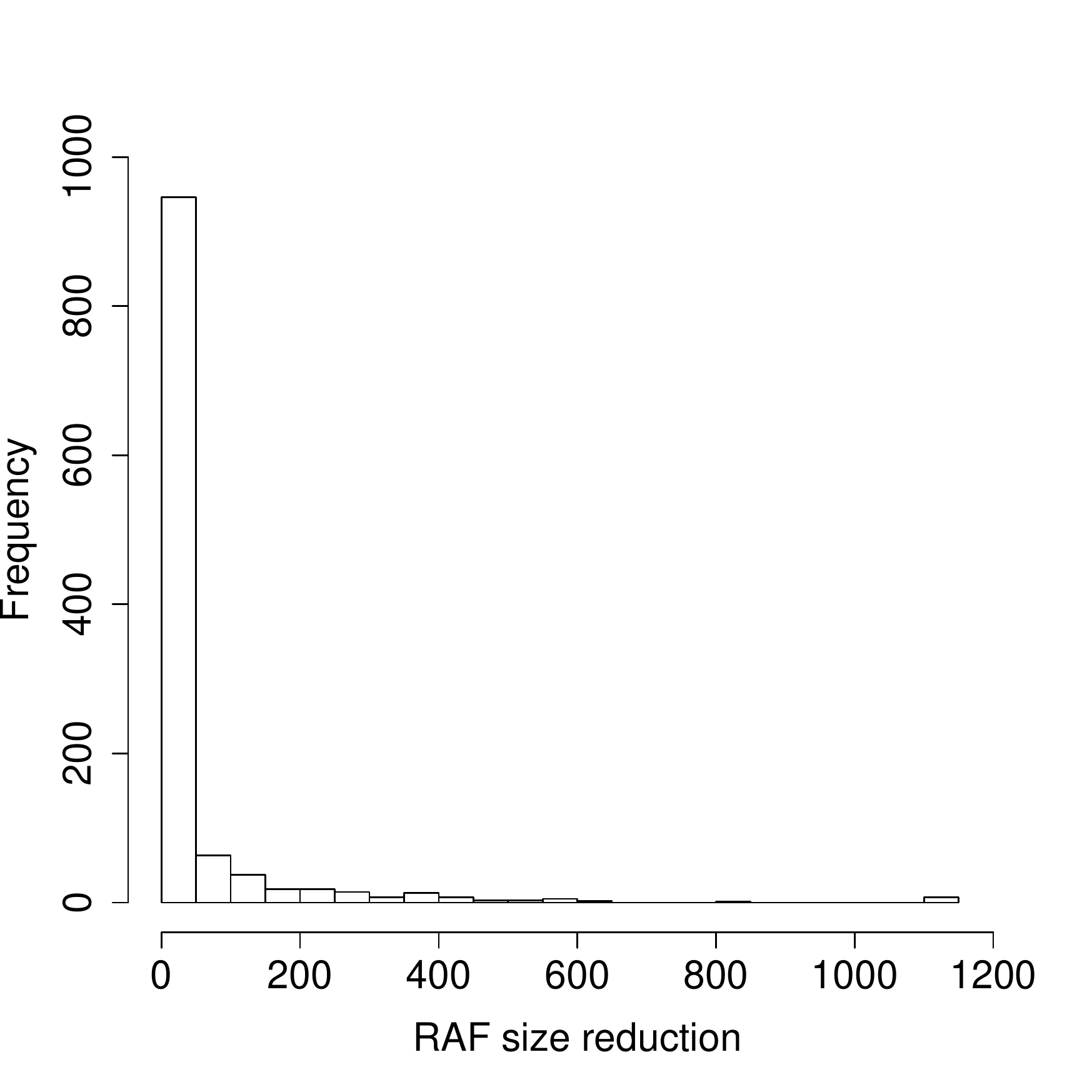}
\caption{Left: The number of goods $M_t$ (solid line) during the last few steps of a single run, and the size of the RAF (dashed line) it contains. Right: A histogram of the number of production functions by which the maxRAF from the final step ($t=196$) decreases upon removal of each production function individually.}
\label{fig:steps}
\end{figure}

For the range of $p_c$ values used so far, a RAF usually only shows up in the final step, when $M_t > 1000$ is reached. And as Figure \ref{fig:RAF_size} shows, they tend to be very large immediately. To see if it is possible that RAFs might first emerge as smaller sets and then grow over time, one single run of the TAP model was done with a larger catalysis probability $p_c = 0.0100$. In this case, a RAF shows up a few time steps before the end of the run. Figure \ref{fig:steps} (left) shows the last 9 steps ($t=188$ to $t=196$) of this particular run, with the number of goods $M_t$ represented by the solid line and the maxRAF size represented by the dashed line.

As the figure shows, at time step $t=190$ a RAF shows up and immediately consists of a large fraction of the full network ($M_{190}=405$, size(RAF) = 337). In just a few more time steps the RAF then grows to encompass virtually the entire network, which it does at the final time step (both being of size 1144).

Finally, the importance of individual production functions is investigated in the RAF that was found in the final step of the single run just presented (with $p_c=0.0100$). Previously, a version of the TAP model that also includes a death rate $\mu$ was investigated \cite{Steel:20}. In other words, goods can ``die'' with a given probability, meaning they cannot be used anymore as a parent to create new goods. Here, a death rate of $\mu=0.0$ was used in all experiments. However, the impact of such ``death'' events can still be inferred by removing each individual production function from the maxRAF, and then applying the RAF algorithm again. This will obviously result in a smaller RAF. In some cases, the reduction may be only by one production function (the one just removed). But in some cases this affects other production functions in the RAF as well (as they may now have lost their catalyst, or an input), and the reduction in RAF size could be much larger. The initially removed production function is then put back into the original RAF, and the next production function is removed and its impact on the size of the remaining RAF is measured, and so on for all individual production functions.

Figure \ref{fig:steps} (right) shows a histogram of the RAF size reductions (measured in number of production functions) caused by these individual removals. As the histogram shows, the far majority of production functions have little impact on the RAF size. However, there are also a few that have a very large impact, sometimes even reducing the RAF size to almost nothing. But overall, the RAF seems to be quite robust and resilient against random removal of production functions. This property was also observed in the binary polymer model.

\subsection{Irreducible RAFs}

As mentioned above, the $M_0$ initial goods (i.e., the food set) were allowed to be catalysts. This often gives rise to a small number of production functions where an initial good is transformed into a new one, also catalyzed by one of the initial goods. Such a production function by itself forms a (sub)RAF of size one. When searching for the smallest RAF subsets (i.e., irreducible RAFs, or irrRAFs) in a larger (max)RAF, it is consequently always these irrRAFs of size one that are found.

To get a better idea of the presence (and sizes) of irrRAFs, one simulation run was performed where the initial goods were {\it not} allowed to be catalysts. A catalysis probability $p_c=0.0044$ was used, which still gives a probability of finding a RAF of close to 0.5 over a large set of runs. For one particular run, which resulted in a (max)RAF of 1101 production functions, a random sample (of size 100) of irrRAFs was obtained.

\begin{figure}[htb]
\centering
\includegraphics[scale=0.4]{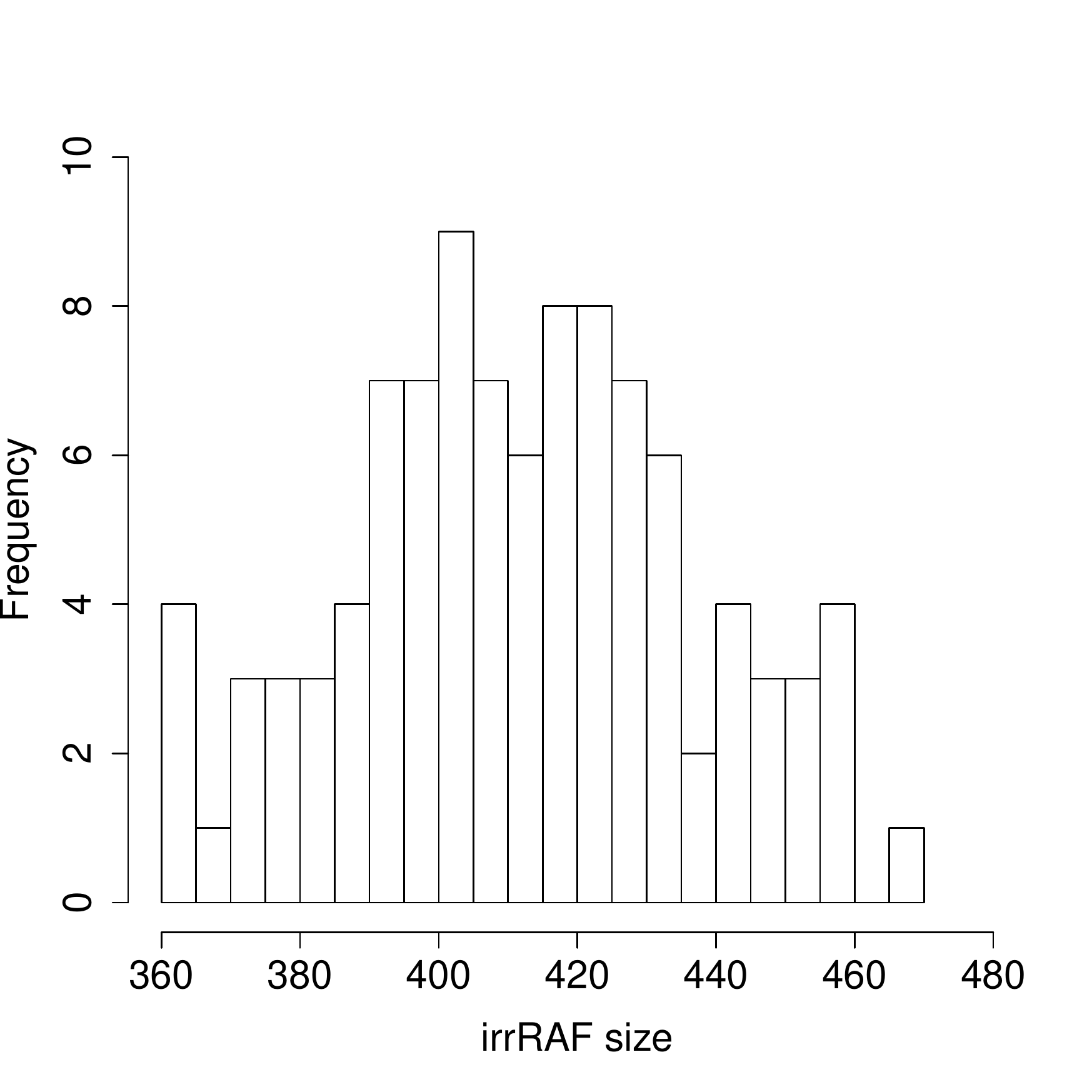}
\caption{A histogram of irrRAF sizes.}
\label{fig:irrRAFs}
\end{figure}

The distribution of irrRAF sizes from this sample is presented in Figure \ref{fig:irrRAFs}. These sizes range from about 360 to 470, with an average of 412 (i.e., slightly more than one-third of the maxRAF). Although all 100 irrRAFs in the sample are different, they have an average overlap of close to 80\%, meaning that for any (arbitrary) pair of irrRAFs from the sample, about 80\% of their production functions are the same.

\section{Conclusions}

In the true spirit of combinatorial innovation, two different views of an economy, one of technological evolution as a process of combinatorial innovation and one of an autocatalytic set of production functions, were combined into one single modeling framework. Using the TAP and RAF models, we have shown that they are compatible, and that autocatalytic (RAF) sets are highly likely to emerge in production function networks that result from a process of combinatorial innovation (TAP).

Depending on the probability $p_c$ that any good can catalyze any production function, a production function network needs to be of a certain minimum size to have a RAF set emerge. But when it does, such a RAF set likely consists of almost the entire production function network. However, smaller autocatalytic subsets (irrRAFs) exist within the larger RAF, on average about one-third the size of the full network.

Many of the results obtained here are similar to those from earlier studies on RAF sets in a simple model of polymer chemistry, which is both reassuring and encouraging. However, there are some subtle differences, mostly resulting from the difference in the ratio between the number of reactions (production functions) and the number of molecules (goods). In the binary polymer model, this ratio increases linearly with the length of the largest polymers, whereas in the TAP model this ratio is constant. In other words, in the polymer model there are many reactions that can create a given molecule, whereas in the TAP model there is always only one production function that creates a given good.

Of course there are many possible variations of the basic model as presented and analyzed here. For example, a power law distribution in catalysis could be used, where most goods do not catalyze anything and a few goods catalyze many production functions. Or an economy could be ``partitioned'' into specialized sub-markets. For example, one market could specialize on everything to do with forestry, and another market on everything related to IT, but goods can act as catalysts both within and between markets (e.g., computers are helpful in any market, and wooden desks are needed for informaticians to work on). However, results on the binary polymer model have shown that such variations of the basic model often lead to very similar results, at least qualitatively. Moreover, sometimes the quantitative differences can be predicted mathematically from the basic model \cite{Hordijk:17}. We expect the same to be true for the TAP model.

The initial results presented here are promising, and warrant further investigation. They strongly support earlier claims that the economy can indeed be seen as an autocatalytic set. Furthermore, they provide deeper insights into how resilient such an economic RAF is, by measuring the impact on the RAF size of the removal of individual production functions. Or they can provide a measure of redundancy, in terms of the amount of overlap between different irrRAFs. We hope that these initial results will stimulate further research in this direction.

% Cite and discuss the Enquist (2011) paper!

\backmatter

\bmhead{Acknowledgments}

The authors thank Mike Steel for helpful comments for improving an earlier version of this article. No funding was received to perform this work, and no conflict of interest is declared.

\bibliography{TAP_RAF}


\begin{thebibliography}{22}
\ifx \bisbn   \undefined \def \bisbn  #1{ISBN #1}\fi
\ifx \binits  \undefined \def \binits#1{#1}\fi
\ifx \bauthor  \undefined \def \bauthor#1{#1}\fi
\ifx \batitle  \undefined \def \batitle#1{#1}\fi
\ifx \bjtitle  \undefined \def \bjtitle#1{#1}\fi
\ifx \bvolume  \undefined \def \bvolume#1{\textbf{#1}}\fi
\ifx \byear  \undefined \def \byear#1{#1}\fi
\ifx \bissue  \undefined \def \bissue#1{#1}\fi
\ifx \bfpage  \undefined \def \bfpage#1{#1}\fi
\ifx \blpage  \undefined \def \blpage #1{#1}\fi
\ifx \burl  \undefined \def \burl#1{\textsf{#1}}\fi
\ifx \doiurl  \undefined \def \doiurl#1{\url{https://doi.org/#1}}\fi
\ifx \betal  \undefined \def \betal{\textit{et al.}}\fi
\ifx \binstitute  \undefined \def \binstitute#1{#1}\fi
\ifx \binstitutionaled  \undefined \def \binstitutionaled#1{#1}\fi
\ifx \bctitle  \undefined \def \bctitle#1{#1}\fi
\ifx \beditor  \undefined \def \beditor#1{#1}\fi
\ifx \bpublisher  \undefined \def \bpublisher#1{#1}\fi
\ifx \bbtitle  \undefined \def \bbtitle#1{#1}\fi
\ifx \bedition  \undefined \def \bedition#1{#1}\fi
\ifx \bseriesno  \undefined \def \bseriesno#1{#1}\fi
\ifx \blocation  \undefined \def \blocation#1{#1}\fi
\ifx \bsertitle  \undefined \def \bsertitle#1{#1}\fi
\ifx \bsnm \undefined \def \bsnm#1{#1}\fi
\ifx \bsuffix \undefined \def \bsuffix#1{#1}\fi
\ifx \bparticle \undefined \def \bparticle#1{#1}\fi
\ifx \barticle \undefined \def \barticle#1{#1}\fi
\bibcommenthead
\ifx \bconfdate \undefined \def \bconfdate #1{#1}\fi
\ifx \botherref \undefined \def \botherref #1{#1}\fi
\ifx \url \undefined \def \url#1{\textsf{#1}}\fi
\ifx \bchapter \undefined \def \bchapter#1{#1}\fi
\ifx \bbook \undefined \def \bbook#1{#1}\fi
\ifx \bcomment \undefined \def \bcomment#1{#1}\fi
\ifx \oauthor \undefined \def \oauthor#1{#1}\fi
\ifx \citeauthoryear \undefined \def \citeauthoryear#1{#1}\fi
\ifx \endbibitem  \undefined \def \endbibitem {}\fi
\ifx \bconflocation  \undefined \def \bconflocation#1{#1}\fi
\ifx \arxivurl  \undefined \def \arxivurl#1{\textsf{#1}}\fi
\csname PreBibitemsHook\endcsname

%%% 1
\bibitem{Arthur:09}
\begin{bbook}
\bauthor{\bsnm{Arthur}, \binits{W.B.}}:
\bbtitle{The Nature of Technology}.
\bpublisher{Free Press},
\blocation{New York, NY, USA}
(\byear{2009})
\end{bbook}
\endbibitem

%%% 2
\bibitem{Enquist:11}
\begin{barticle}
\bauthor{\bsnm{Enquist}, \binits{M.}},
\bauthor{\bsnm{Ghirlanda}, \binits{S.}},
\bauthor{\bsnm{Eriksson}, \binits{K.}}:
\batitle{Modelling the evolution and diversity of cumulative culture}.
\bjtitle{Philosophical Transactions of the Royal Society B}
\bvolume{366},
\bfpage{412}--\blpage{423}
(\byear{2011})
\end{barticle}
\endbibitem

%%% 3
\bibitem{Steel:20}
\begin{barticle}
\bauthor{\bsnm{Steel}, \binits{M.}},
\bauthor{\bsnm{Hordijk}, \binits{W.}},
\bauthor{\bsnm{Kauffman}, \binits{S.A.}}:
\batitle{Dynamics of a birth-death process based on combinatorial innovation}.
\bjtitle{Journal of Theoretical Biology}
\bvolume{491},
\bfpage{110187}
(\byear{2020})
\end{barticle}
\endbibitem

%%% 4
\bibitem{Roser:13}
\begin{botherref}
\oauthor{\bsnm{Roser}, \binits{M.}}:
Economic growth.
Our World in Data
(2013).
(https://ourworldindata.org/economic-growth)
\end{botherref}
\endbibitem

%%% 5
\bibitem{Karakas:19}
\begin{botherref}
\oauthor{\bsnm{Karakas}, \binits{A.D.}}:
Destiny of the 'hockey stick' economic growth.
The Economics Review
(2019).
(https://theeconreview.com/2019/09/12/destiny-of-the-hockey-stick-economic-growth/)
\end{botherref}
\endbibitem

%%% 6
\bibitem{Kauffman:11}
\begin{botherref}
\oauthor{\bsnm{Kauffman}, \binits{S.A.}}:
Economics and the collectively autocatalytic structure of the real economy.
NPR 13.7 Cosmos \& Culture
(2011).
(https://www.npr.org/sections/13.7/2011/11/21/142594308/economics-and-the-collectively-autocatalytic-structure-of-the-real-economy)
\end{botherref}
\endbibitem

%%% 7
\bibitem{Hordijk:13}
\begin{barticle}
\bauthor{\bsnm{Hordijk}, \binits{W.}}:
\batitle{Autocatalytic sets: From the origin of life to the economy}.
\bjtitle{BioScience}
\bvolume{63}(\bissue{11}),
\bfpage{877}--\blpage{881}
(\byear{2013})
\end{barticle}
\endbibitem

%%% 8
\bibitem{Gatti:20}
\begin{barticle}
\bauthor{\bsnm{{Cazzolla Gatti}}, \binits{R.}},
\bauthor{\bsnm{Koppl}, \binits{R.}},
\bauthor{\bsnm{Fath}, \binits{B.D.}},
\bauthor{\bsnm{Kauffman}, \binits{S.}},
\bauthor{\bsnm{Hordijk}, \binits{W.}},
\bauthor{\bsnm{Ulanowicz}, \binits{R.E.}}:
\batitle{On the emergence of ecological and economic niches}.
\bjtitle{Journal of Bioeconomics}
\bvolume{22},
\bfpage{99}--\blpage{127}
(\byear{2020})
\end{barticle}
\endbibitem

%%% 9
\bibitem{Koppl:22}
\begin{botherref}
\oauthor{\bsnm{Koppl}, \binits{R.}},
\oauthor{\bsnm{{Cazzolla Gatti}}, \binits{R.}},
\oauthor{\bsnm{Deveraux}, \binits{A.}},
\oauthor{\bsnm{Fath}, \binits{B.D.}},
\oauthor{\bsnm{Herriot}, \binits{J.}},
\oauthor{\bsnm{Hordijk}, \binits{W.}},
\oauthor{\bsnm{Kauffman}, \binits{S.}},
\oauthor{\bsnm{Ulanowicz}, \binits{R.E.}},
\oauthor{\bsnm{Valverde}, \binits{S.}}:
Explaining technology.
Cambridge Elements Series on Evolutionary Economics
(2022).
(Under review)
\end{botherref}
\endbibitem

%%% 10
\bibitem{Kauffman:71}
\begin{barticle}
\bauthor{\bsnm{Kauffman}, \binits{S.A.}}:
\batitle{Cellular homeostasis, epigenesis and replication in randomly
  aggregated macromolecular systems}.
\bjtitle{Journal of Cybernetics}
\bvolume{1}(\bissue{1}),
\bfpage{71}--\blpage{96}
(\byear{1971})
\end{barticle}
\endbibitem

%%% 11
\bibitem{Kauffman:86}
\begin{barticle}
\bauthor{\bsnm{Kauffman}, \binits{S.A.}}:
\batitle{Autocatalytic sets of proteins}.
\bjtitle{Journal of Theoretical Biology}
\bvolume{119},
\bfpage{1}--\blpage{24}
(\byear{1986})
\end{barticle}
\endbibitem

%%% 12
\bibitem{Hordijk:04}
\begin{barticle}
\bauthor{\bsnm{Hordijk}, \binits{W.}},
\bauthor{\bsnm{Steel}, \binits{M.}}:
\batitle{Detecting autocatalytic, self-sustaining sets in chemical reaction
  systems}.
\bjtitle{Journal of Theoretical Biology}
\bvolume{227}(\bissue{4}),
\bfpage{451}--\blpage{461}
(\byear{2004})
\end{barticle}
\endbibitem

%%% 13
\bibitem{Hordijk:17}
\begin{barticle}
\bauthor{\bsnm{Hordijk}, \binits{W.}},
\bauthor{\bsnm{Steel}, \binits{M.}}:
\batitle{Chasing the tail: The emergence of autocatalytic networks}.
\bjtitle{BioSystems}
\bvolume{152},
\bfpage{1}--\blpage{10}
(\byear{2017})
\end{barticle}
\endbibitem

%%% 14
\bibitem{Ashkenasy:04}
\begin{barticle}
\bauthor{\bsnm{Ashkenasy}, \binits{G.}},
\bauthor{\bsnm{Jegasia}, \binits{R.}},
\bauthor{\bsnm{Yadav}, \binits{M.}},
\bauthor{\bsnm{Ghadiri}, \binits{M.R.}}:
\batitle{Design of a directed molecular network}.
\bjtitle{PNAS}
\bvolume{101}(\bissue{30}),
\bfpage{10872}--\blpage{10877}
(\byear{2004})
\end{barticle}
\endbibitem

%%% 15
\bibitem{Vaidya:12}
\begin{barticle}
\bauthor{\bsnm{Vaidya}, \binits{N.}},
\bauthor{\bsnm{Manapat}, \binits{M.L.}},
\bauthor{\bsnm{Chen}, \binits{I.A.}},
\bauthor{\bsnm{Xulvi-Brunet}, \binits{R.}},
\bauthor{\bsnm{Hayden}, \binits{E.J.}},
\bauthor{\bsnm{Lehman}, \binits{N.}}:
\batitle{Spontaneous network formation among cooperative {RNA} replicators}.
\bjtitle{Nature}
\bvolume{491},
\bfpage{72}--\blpage{77}
(\byear{2012})
\end{barticle}
\endbibitem

%%% 16
\bibitem{Arsene:18}
\begin{barticle}
\bauthor{\bsnm{Ars\`{e}ne}, \binits{S.}},
\bauthor{\bsnm{Ameta}, \binits{S.}},
\bauthor{\bsnm{Lehman}, \binits{N.}},
\bauthor{\bsnm{Griffiths}, \binits{A.D.}},
\bauthor{\bsnm{Nghe}, \binits{P.}}:
\batitle{Coupled catabolism and anabolism in autocatalytic {RNA} sets}.
\bjtitle{Nucleic Acids Research}
\bvolume{46}(\bissue{18}),
\bfpage{9660}--\blpage{9666}
(\byear{2018})
\end{barticle}
\endbibitem

%%% 17
\bibitem{Miras:20}
\begin{barticle}
\bauthor{\bsnm{Miras}, \binits{H.N.}},
\bauthor{\bsnm{Mathis}, \binits{C.}},
\bauthor{\bsnm{Xuan}, \binits{W.}},
\bauthor{\bsnm{Long}, \binits{D.-L.}},
\bauthor{\bsnm{Pow}, \binits{R.}},
\bauthor{\bsnm{Cronin}, \binits{L.}}:
\batitle{Spontaneous formation of autocatalytic sets with self-replicating
  inorganic metal oxide clusters}.
\bjtitle{PNAS}
\bvolume{117}(\bissue{20}),
\bfpage{10699}--\blpage{10705}
(\byear{2020})
\end{barticle}
\endbibitem

%%% 18
\bibitem{Sousa:15}
\begin{barticle}
\bauthor{\bsnm{Sousa}, \binits{F.L.}},
\bauthor{\bsnm{Hordijk}, \binits{W.}},
\bauthor{\bsnm{Steel}, \binits{M.}},
\bauthor{\bsnm{Martin}, \binits{W.F.}}:
\batitle{Autocatalytic sets in {{\it E. coli}} metabolism}.
\bjtitle{Journal of Systems Chemistry}
\bvolume{6},
\bfpage{4}
(\byear{2015})
\end{barticle}
\endbibitem

%%% 19
\bibitem{Xavier:20}
\begin{barticle}
\bauthor{\bsnm{Xavier}, \binits{J.C.}},
\bauthor{\bsnm{Hordijk}, \binits{W.}},
\bauthor{\bsnm{Kauffman}, \binits{S.A.}},
\bauthor{\bsnm{Steel}, \binits{M.}},
\bauthor{\bsnm{Martin}, \binits{W.F.}}:
\batitle{Autocatalytic chemical networks at the origin of metabolism}.
\bjtitle{Proceedings of the Royal Society B}
\bvolume{287},
\bfpage{20192377}
(\byear{2020})
\end{barticle}
\endbibitem

%%% 20
\bibitem{Kauffman:21}
\begin{botherref}
\oauthor{\bsnm{Kauffman}, \binits{S.A.}},
\oauthor{\bsnm{Roli}, \binits{A.}}:
The third transition in science: Beyond {N}ewton and quantum mechanics.
arXiv
(2021).
(https://arxiv.org/abs/2106.15271)
\end{botherref}
\endbibitem

%%% 21
\bibitem{Temkin:96}
\begin{bbook}
\bauthor{\bsnm{Temkin}, \binits{O.N.}},
\bauthor{\bsnm{Zeigarnik}, \binits{A.V.}},
\bauthor{\bsnm{Bonchev}, \binits{D.}}:
\bbtitle{Chemical Reaction Networks: A Graph-Theoretical Approach}.
\bpublisher{CRC Press},
\blocation{Boca Raton, FL, USA}
(\byear{1996})
\end{bbook}
\endbibitem

%%% 22
\bibitem{Hordijk:15}
\begin{barticle}
\bauthor{\bsnm{Hordijk}, \binits{W.}},
\bauthor{\bsnm{Smith}, \binits{J.I.}},
\bauthor{\bsnm{Steel}, \binits{M.}}:
\batitle{Algorithms for detecting and analysing autocatalytic sets}.
\bjtitle{Algorithms for Molecular Biology}
\bvolume{10},
\bfpage{15}
(\byear{2015})
\end{barticle}
\endbibitem

\end{thebibliography}

\end{document}